\documentclass[a4paper,12pt]{article}

\usepackage[utf8]{inputenc}
\usepackage{epsfig}
\usepackage{graphicx}
\usepackage{amsmath}
\usepackage{amssymb}
\usepackage{cmap}
\usepackage{bm}
\usepackage{hyperref}
\usepackage{array}
\usepackage{epsfig}
\usepackage{array}
\usepackage{color}
\usepackage{latexsym}
\usepackage{xypic}
\usepackage{amscd}
\usepackage{lipsum}

\usepackage{svg}

\usepackage{cite} 

\usepackage[left=0.8in, right=0.8in, top=0.8in, bottom=0.8in]{geometry}

\newcommand{\PRLsep}{\noindent\makebox[\linewidth]{\resizebox{0.4\linewidth}{1.2pt}{$\bullet$}}\bigskip}

\newcommand{\version}{\sf draft~for~\textbf{arXiv:~v2}\ ---\ \today}

\title{Nonperturbative quark-gluon thermodynamics at finite density\thanks{\version}}

\author{M.A.~Andreichikov$^{a,}$\thanks{andreichicov@mail.ru},\,\, M.S.~Lukashov$^{a,b,}$\thanks{lukashov@phystech.edu}\,\, and Yu.A.~Simonov$^{a,}$\thanks{simonov@itep.ru}\, \\
\\
$^a$ \small{\em Alikhanov Institute for Theoretical and Experimental Physics,}\\
\small{\em B. Cheremushkinskya 25, 117218 Moscow, Russia}\vspace{0.25cm}\\
$^b$ \small{\em Moscow Institute of Physics and Technology,}\\
\small{\em Institutskiy per. 9, 141700 Dolgoprudny, Moscow Region, Russia}
}\bigskip
\date{\today}

\newcommand{\be}{\begin{equation}}
\newcommand{\ee}{\end{equation}}

\def\la{\mathrel{\mathpalette\fun <}}

\def\fun#1#2{\lower3.6pt\vbox{\baselineskip0pt\lineskip.9pt
\ialign{$\mathsurround=0pt#1\hfil ##\hfil$\crcr#2\crcr\sim\crcr}}}

\newcommand{{\SD}}{\rm SD}

\newcommand{{\Mc}}{\mathcal{M}}

\newcommand{\lan}{\langle}
\newcommand{\ran}{\rangle}

\begin{document}

\maketitle
\begin{abstract}
 Thermodynamics of the quark-gluon plasma at finite density is studied in the
 framework of the Field Correlator Method, where thermodynamical effects of
 Polyakov loops and colormagnetic confinement are taken into account.  Having
 found good agreement with numerical lattice data for zero density, we calculate pressure $P(T,\,\mu)$ for $0<\mu<400$ MeV and $150 <T< 1000$ MeV. For the first time the explicit intergral  form is found in this region,demonstrating analytic structure in the complex $\mu$ plane. The resulting multiple complex branch points are found at the Roberge-Weiss values of $\operatorname{Im}\,\mu$, with $\operatorname{Re}\,\mu$ defined by the values of Polyakov lines and colormagnetic confinement.

\end{abstract}

\newpage

\section{Introduction }
The  new phenomenon of the  quark-gluon plasma (QGP) was predicted in
\cite{1,2,3}, and its properties were  measured soon on the lattice
\cite{4,5,6}. Nowadays the QGP as an important stage in the heavy-ion
collisions is widely recognized, see e.g. \cite{7a,7b,8a,8b,9}.

The recent accurate lattice measurements both in  SU(3) \cite{10,11a,11b} and in
 realistic $2+1$ QCD theory \cite{12,13} have revealed the nontrivial
character of thermodynamics for $T<600$ MeV, which includes strong
nonperturbative \textit{(np)} interaction during the temperature transition region and
beyond it.

At larger $T$, $T>600$ MeV, one can hope to rely   on the thermal perturbation
theory (HTL) \cite{14,15,16}, however the infrared  Linde problem implies an
important role of \textit{np}  interaction also here \cite{17a,17b}.

Therefore the \textit{np} thermodynamics seems to be unavoidable in the whole $T$
region.

The corresponding \textit{np}  approach, based on the vacuum fields, the Field
Correlator Method (FCM), was originally formulated in \cite{18a,18b,18c,18d,18e,18f,18g}, where the
deconfining phase transition was associated with the vanishing of the confining
correlator $D^E $, and later in \cite{19a,19b} it was shown, that another \textit{np}
correlator, $D^E_1$, is responsible for the  dynamics of Polyakov loops.
Moreover, the final form of the \textit{np} thermodynamics in FCM was formulated in
\cite{19a,19b} and compared with existing lattice data. At that time in the region
beyond $T_c$ only Polyakov lines have been taken into account, however  a
reasonable agreement within (20-25)\%  with  the lattice data was achieved.

Recently another important ingredient of the \textit{np} interaction in the region
$T>T_c$ was taken into account  in addition to Polyakov lines -- the Color
Magnetic (CM) confinement is given by the spatial projection of the Wilson loop, and the \textit{np} theory based on FCM
for zero density (chemical potential) was finally formulated. In \cite{20,21}
this theory was fully investigated in the case of SU(3) and compared to the
accurate lattice data \cite{10}, showing a good agreement for $p(T), I(T)$ both
below and above $T_c$, and also in the character of transition. It is
important, that a new effect was used in the subcritical region -- the gradual
vanishing of the confinement (the string tension $\sigma$) when $T$ approaches
$T_c$ from below.

This fact was found earlier on the lattice \cite{22a,22b,22c,22d}, and the use of this makes
it unnecessary to exploit the Hagedorn string spectra.

In the quark-gluon case of $2+1$ QCD the same type of the \textit{np} approach (the
account of Polyakov lines and CM confinement) was done for zero density and in
the deconfined region in \cite{23}. The  resulting thermodynamic  potentials in
\cite{23} are in a good agreement with the accurate lattice data \cite{12,13}
in the region 150 MeV $<T< 1000$ MeV, which implies, that the main part of
dynamics is correctly taken into account.  The confining region $T\la 150$ MeV
was not treated in \cite{23} and is in progress.

Meanwhile the region of nonzero density (chemical potential $\mu$)  is of the
outmost importance. Indeed, the existing and planned experiments badly need the
corresponding theoretical calculations of the QGP properties at nonzero $\mu$,
whereas the lattice data are not directly available in  this region. One
particular example of an indirect information, provided by lattice data, is
quark numbers susceptibilities $\chi_n^X$, which   exist for long time
\cite{24},  see  \cite{25,26} for recent data. These data are important for
comparison with experimental results in the freeze-out region \cite{7a,7b,8a,8b,9}.
It is a basic feature of our approach,that the nonzero density is easily incorporated
into the formalism,and the analysis of the complex chemical potential can be done in the
whole plane of complex $\mu$.In this way one can find all singularities in this plane
and find their dynamical origin.This can be closely connected to the possible existence
of critical and quasicritical points in $(\mu,\,T)$ plane,and is therefore of utmost importance.

Our  \textit{np} approach to the  case of the finite density was formulated in
\cite{27a,27b}, see also \cite{28} for a review, where in the QGP only the Polyakov
line interaction was taken into account, and the whole temperature transition
curve in the $\mu - T$ plane was found.

In addition the curvature of $T_c(\mu)$  was found in \cite{27a,27b},
$\kappa_2=0.0110(3)$, which is  in  a good agreement with existing  lattice and
freeze-out data, see Fig. 1 in  \cite{29}.

However, the CM confinement was not taken into account in \cite{27a,27b}, and the
experience of our latest calculations in SU(3) and QCD shows, that  it is
important and can seriously improve the accuracy of the results.

The purpose of the present paper is to incorporate in our calculations of \textit{np}
thermodynamics for finite density the effects CM  confinement and to produce the
function of pressure $p(T,\mu)$, in the temperature interval $0.2<T<1.0$ GeV.
It is also interesting to investigate the properties of $p(T,\mu)$ in the whole complex plane of $\mu$
and to compare with  known lattice information.

The paper is organized as follows. In the next section the main equations of
our method are presented. In section 3 the properties of the thermodynamical
potentials  are discussed,and in section 4 the case of an arbitrary CM interaction is treated,
 while the section 5 contains numerical results and the section 6 is devoted to the
discussion and an outlook.%

\section{Thermodynamic potentials  of quark-gluon plasma at finite density}

We are using below the same gauge and relativistic invariant formalism, based
on the  path  integral  formalism, which was  formulated in \cite{18a,18b,18c,18d,18e,18f,18g,19a,19b}, and
exploited in the  $SU(3)$ case in \cite{23}. The basic interaction of a quark
or  a gluon can be expressed via world lines affected by the vacuum fields and
finally  written in the form of Wilson--lines and loops. One can consider \cite{18c,19a}
as a detailed review of FCM technique applied to the np quark-gluon thermodynamics at finite temperature and density.
It is  essential that in the  deconfined phase two basic interactions define
the quark and gluon  dynamics: the colorelectric (CE) one, contained in the
Polyakov line $L(T)$, and the colormagnetic (CM) interaction in the spatial
projection on the Wilson loop. The CE part is expressed via the \textit{np} part of the
CE field correlator $D^E_1 (\tau)$, while perturbative part of $D_1$ yields
color Coulomb potential. The CM part is defined by the CM field correlator
$D^H(z)$, yielding the  spatial string  tension $\sigma _s (T)= \frac12 \int
D^H (z) d^2 z$. As it was shown  within FCM \cite{17b} $\sigma_s (T) = O(T^2)$
and is important in the whole region $T\geq T_c$.

Using the $T$ dependent path integral (world line) formalism one can express
thermodynamic potentials via the Wilson loop integral, e.g. for the gluon
pressure one has \cite{17b,21}

\be P_{gl} = 2 (N^2_c -1) \int^\infty_0 \frac{ds}{s} \sum_{n=1,2...}
G^{(n)}{(s)},\label{1*}\ee where we use Feynman-Fock-Schwinger (FSF) formalism with Schwinger proper time $s$, and $G^{(n)} (s)$ are the winding (Matsubara) path
integrals \be G^{(n)} (s) = \int (Dz)^w_{on} e^{-K} \lan \hat{tr}_a
W(C_n)\ran,\label{2*}\ee
where $K = \frac{1}{4}\int_0^s\left(\frac{dz_{\mu}(\tau)}{d\tau}\right)^2d\tau$ according to the FSF method and $W(C_n)$ is the Wilson loop defined for the gluon path $C_n$, which  has
both temporal (i4) and spacial projections (ij). It is important, that the  CE
and CM field strengths in $T >T_c$ region correlate very weakly during the gauge-invariant field correlator in adjoint representation $\lan E_i (x) B_k (y) \Phi(x,y)\ran
\approx 0$ (see \cite{19a,19b} for a discussion of this point) and therefore both
CE and CM projections of the $ \hat{tr}_a W(C_n)$ can be factorized as shown in
\cite{21}

\be \lan tr_a W(C_n)\ran = L_{\rm adj}^{(n)} (T) \lan W_3\ran.\label{3*}\ee

Inserting (\ref{3*}) in (\ref{2*}), one can integrate out the $z_4$ part of the
path integral $(Dz)^w_{on}=(Dz_4)^w_{on} D^3 z$, and write the result as

\be G^{(n)} (s) =  G_4^{(n)} (s) G_3 (s) ;~~ G_4^{(n)} (s)= \int (Dz)^w_{on}
e^{-K}  L_{adj}^{(n)} = \frac{1}{2\sqrt{4\pi s}} e^{-n^2/4T^2s} L_{adj}^{(n)}.
\label{4*}\ee

This basic factorization holds also for quarks and will be used below for both
quarks and gluons.\medskip

 {\sc a) gluons}\smallskip

 Following the previous  discussion we start with the gluon pressure and write it in the same form, as was written
in \cite{19a,19b,28,21}, where gluon is moving in the vacuum field,  which creates
 both Polyakov loops (via interaction $V_1 (\infty, T)$) and
confinement for the spatial loops

 \be P_{gl} = \frac{N_c^2-1}{\sqrt{4\pi}} \int^\infty_0 \frac{ds}{s^{3/2}}
G_3 (s) \sum_{n=0,1,2...} e^{-\frac{n^2}{4T^2s} } L_{\rm
adj}^{(n)}.\label{1}\ee

Here $ G_3 (s)$ is the gluon Green's function in the 3d spatial projection. It
can be written in the free case as $G_3^{(0)}  (s)=\frac{1}{(4\pi s)^{3/2}}$,
and when the interaction in the loop has the form of  an oscillator, it can be
written as \cite{21}

\be G_3^{osc} (s) = \frac{1}{(4\pi s)^{3/2} } \frac{(M_{0}^2\,s)}{\sinh (M_{0}^2\,s)},\label{2}\ee where $M_0$ is connected to the oscillator parameters. In
\cite{21} the following approximate form was suggested in the realistic case of
the linear confinement \be G_3^{lin} (s) = \frac{1}{(4\pi s)^{3/2}}
 \sqrt{\frac{(M^2_{\rm adj}\,
 s)}{\sinh (M_{\rm adj}^2\,s)}} .\label{3}\ee

 The form (\ref{3}) with $M_{\rm adj} \cong 2M_D$ ($M_D$  is the gluon Debye
 mass) used in \cite{21} provides the pressure $p(T)$ and trace anomaly $I(T)$
 in good agreement with lattice data \cite{10}. As shown in \cite{19a,19b,28}, $L_{\rm
 adj}^{(n)} \cong (L_{\rm adj} (T))^n$ for $T\la 1 $ GeV, and $L_{\rm adj} (T)
 = \exp \left(- \frac{9V_1 (\infty,T)}{8T}\right)$. As shown in \cite{21}, the
 resulting $L_{\rm adj} (T)$, which is  close to the lattice measurement values, yields
 realistic thermodynamic potentials.\medskip

 {\sc b) quarks}\smallskip

 For quarks one can write, following \cite{19a,19b,27a,27b} the same form as in
 (\ref{1}), but augmented by the quark mass term $e^{-m^2_f s}$ and the density
 term $\cosh \frac{\mu n}{T}$.

\be  P_{q}= \sum_{m_q} P_q^{(f)}, ~~ P_{q}^{(f)} = \frac{4N_c }{\sqrt{4\pi}}
\int^\infty_0 \frac{ds}{s^{3/2}}
 e^{-m^2_fs}
S_3 (s) \sum_{n=  1,2,..}(-)^{ n+1 } e^{-\frac{n^2}{4T^2s} } \cosh\left(
\frac{\mu n}{T}\right) L_{f}^{ (n)  }.\label{4}\ee

Here $S_3(s)$ is, similarly to $G_3(s)$, the 3d quark Green's function, which
can be approximated in the same way as in (\ref{3}), \be S_3^{lin} (s) =
\frac{1}{(4\pi s)^{3/2}}
 \sqrt{\frac{(M^2_{f}\,s)}{\sinh(M_{f}^2\,s)}} ,\label{5}\ee with the  relation $M_{\rm adj}^2 = \frac94
 M^2_f$.

 Eqs.(\ref{1}), (\ref{4}) with definitions (\ref{3}), (\ref{5}) are our basis
 for the analysis and calculations to be done in the following sections.

\section{Properties of thermodynamic potential at finite density}

In what follows we shall be interested in  the full
pressure $P(T, \mu) = P_{gl} (T) + P_q (T, \mu), $ where $P_q (T, \mu)$ is
given in (\ref{4}, (\ref{5}).In addition on the lattice one also considers quark number susceptibilities $\chi_{m,n} (T)$, according to
the standard definitions \be \Delta P (T,\mu)= P(T,\mu)-P(T,\mu =0)\label{6}\ee

\be \frac{ \Delta P (T,\mu)}{T^4} = \sum_{i+j+k=even} \frac{\chi_{ijk}
(T)}{i!j!k!} \hat\mu^{(i)}_u \hat\mu^{(j)}_d\hat\mu_s^{(k)},\label{7}\ee where
$\hat\mu = \frac{\mu}{T}$.

In addition it is useful to obtain the baryon density $n(T,\mu_B)$

\be n(T, \mu_B) = \frac{ \partial \Delta P (T, \mu_B)}{\partial \mu_B}, \mu_B
\cong 3\mu.\label{8}\ee

The analysis of $\chi_{mn} $ in lattice data allows to obtain information on
the possible critical point $T_c(\mu)$.

In our case, since our $ P (T,\mu)$ has a definite analytic form, one can
search for $T_c (\mu)$ explicitly. Indeed, after the integration over $ds$, our
$ P (T, \mu)$ is a sum over $n$, which  diverge at some value of $\mu
= \mu_{cr} (T) \simeq 0.51 \ GeV$.

It is convenient to  represent $P^{(f)}_q$  in (\ref{4})  in the form
 \be
\frac{1}{T^4} P^{(f)}_q= \frac{ N_c}{4\pi^2} \sum^\infty_{n=1}
\frac{(-1)^{n+1}}{n^4} e^{-\frac{nV_1 (\infty, T)}{2T}} \cosh \frac{\mu n}{T}
\cdot \Phi_n (T),\label{10}\ee where $\Phi_n (T)$ is

\be \frac{\Phi_n(T)}{n^4}   = \frac{(4\pi)^{3/2}}{T^4} \int^\infty_0
\frac{ds}{s^{3/2}} e^{-m^2_fs} S_3 (s) e^{-\frac{n^2}{4T^2s}}.\label{a10}\ee
One can use for $S_3(s)$ different forms. E.g., of one expands $S_3(s)$ in
terms, corresponding to the series over bound states with masses $ m_\nu$ and
wave functions $\psi_\nu (x)$,  then one can  write $S_3 (s)$ as in Eq. (34) of
\cite{21}, namely \be S_3 (s) = \frac{1}{\sqrt{\pi s}} \sum_{\nu=0,1,...}
\psi^2_\nu (0) e^{-m^2_\nu s}.\label{a11} \ee In the case, when one represents
the colormagnetic confinement by the oscillator potential, one has \be S_3^{osc} (s)
= \frac{1}{(4\pi s)^{3/2}}\frac{(M^2_0\,s)}{\sinh (M^2_0\,s)}.\label{12}\ee

Finally, when one approximates the linear CM confinement as in (\ref{5}), then
one can write the following form, which apprximates $S_3^{osc}$ and $S_3^{lin}$ with the accuracy of (5-10)\% in the region $M_0^2\,s$

\be S_3^{lin} (s) \cong \frac{1}{(4 \pi s)^{3/2}}\sqrt{ \frac{(M^2\,s)}{\sinh
(M^2s)}} \approx \frac{1}{(4\pi s)^{3/2}}
e^{-\frac{M^2\,s}{4}}.\label{9}\ee
 In this case $\Phi_n(T)$ can be calculated explicitly, namely.

\be \Phi_n (T) = \frac{8n^2 {\bar{M}}^2  }{ T^2} K_2 \left( \frac{\sqrt{
{\bar{M}}^2 }n}{T}\right),\quad {\bar{M}} = \sqrt{ m^2_f + \frac{M^2}{4}}.\label{11}\ee

 Let us introduce the following functions
 \be \xi_1 =\sum_n  \frac{(-)^{n+1}}{n^2} L^n  e^{\pm \frac{\mu n}{T}} K_2
 \left( \frac{\bar M n}{T}\right), \label{21}\ee

 One can use   the  integral representation

 \be K_\nu (z) = \frac{\left( \frac{z}{2}\right)^\nu \Gamma \left(
 \frac{1}{2}\right)}{\Gamma\left(\nu +\frac12\right)} \int^\infty_0 e^{-z\,\cosh{t}}
 \sinh^{z\nu}{t}\,dt, \label{23}\ee
 which allows to sum up all terms in the sum over $n$, namely

 \be \xi^\pm_1 = \frac43 \left(\frac{\bar M}{2T}\right)^2\int^\infty_0
 \frac{u^4 du}{\sqrt{1+u^2}} \frac{1}{1+\exp \left( \frac{\bar M}{T}
 \sqrt{1+u^2} + \frac{V_1}{2T} \mp \frac{\mu}{T} \right)},\label{24}\ee

and the pressure can be
written as \be \frac{1}{T^4} P^{(f)}_q = \frac{2N_c}{\pi^2} \left[ \frac12
\left( \xi^+_1 + \xi_1^- \right)\right]\label{Pxi}\ee  One can see, that (\ref{24}) has no
singularities at $\mu$ real, but $\xi_i$ may get a  singularity  for $\operatorname{Im}\frac{\mu}{T} = \pi$ due to vanishing of  the denominator in (\ref{24}) at $u=0$.
 We can add to $\frac{V_1(\infty, T)}{2T}$  the
complex phase of the Polyakov loop $i\phi$, where $\phi$ can assume $Z(3)$
values $\phi_k = \frac{2\pi}{3}k, ~~ k=0,\pm 1$ and the factor in the exponent in (\ref{24}) at $u=0$ has the form \be a_\pm = \exp \left(\pm
\frac{\mu_R +i \mu_I}{T} + i \frac{2\pi}{3} k + \frac{V_1 (\infty, T)}{2T} +
\frac{\bar M(T)}{T}\right).\label{A2}\ee Note, that for the prefactor of the exponent in (\ref{24})
be equal to $(-1)$ the imaginary part of $\mu$ should be equal  \be \frac{\mu_I}{T} = \frac{\pi}{3}
(2n+1), ~~ n= 0,\pm 1,\pm 2,...\label{A3}\ee

These are exactly the Roberge-Weiss values \cite{30}.

This situation  may explain the appearance of the Roberge-Weiss singularities
\cite{30}, see \cite{31}, \cite{32} for a physical and numerical analysis.

As a result, in the normal situation with  $\mu$ and $L_f$ real the singularity
is absent, implying the absence of the critical point $T_c (\mu)$. This result
is in  line with the lattice analysis in \cite{25,26,33}, where no sign of
$T_c$ was observed in  quark numbers susceptibilities.  Note, however, that our
conclusion refers to the  purely \textit{np} contribution,  where the perturbative
gluon and quark exchanges are absent.  Moreover, we have not taken into account
a possible density modification of the  vacuum  averages, in particular of the
confinement parameters.

\section{The case of  arbitrary CM interaction}

In  the general case of the  CM interaction, which  produces in 3d  the
spectrum with eigenvalues $m^2_\nu$ and eigenfunctions $\psi_\nu (\rho)$, one
can write as in (\ref{a11})

\be S_3 (s) = \frac{1}{\sqrt{\pi s}} \sum_{\nu=0,1,..} \psi^2_\nu (0)
e^{-m^2_\nu s}\label{a1}\ee

\begin{figure}[htb]
\vspace{0.5cm}
\setlength{\unitlength}{1.0cm}
\begin{center}
\vspace{-0.1cm}
\center{\includegraphics[width=0.8\linewidth]{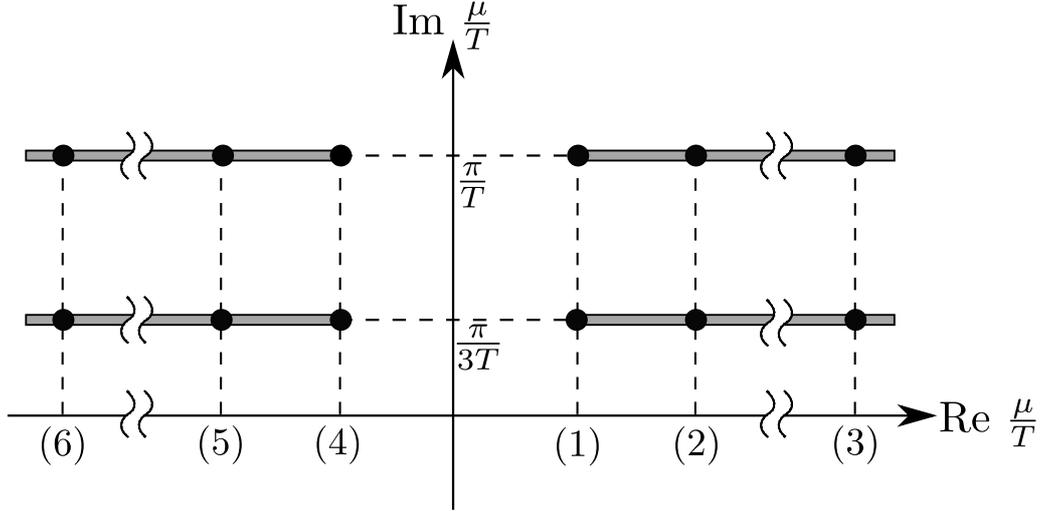}}
\vspace{-0.1cm}
\caption{Roberge-Weiss singular points and cuts in the complex plane of $\mu$. Points 1,2,3,4,5, and 6 are, respectively, $\left(\frac{V_1}{2T}\!+\!\frac{M_0}{T}\right),\,\left(\frac{V_1}{2T}\!+\!\frac{M_0\sqrt{3}}{T}\right),\,\left(\frac{V_1}{2T}\!+\!\frac{M_{\nu}}{T}\right),\,-\left(\frac{V_1}{2T}\!+\!\frac{M_0}{T}\right),\,-\left(\frac{V_1}{2T}\!+\!\frac{M_0\sqrt{3}}{T}\right)$, and $-\left(\frac{V_1}{2T}\!+\!\frac{M_{\nu}}{T}\right)$. In the lower half plane the points are mirror-reflected of the axis $\operatorname{Re}\,(\mu/T)$.}
\label{fig:01}
\end{center}
\end{figure}\smallskip

For $\Phi_n (T)$ in (\ref{10}) one obtains

\be \frac{\Phi_n(T)}{n^4}   = \frac{ 8\pi }{T^4} \sum_\nu
\psi^2_\nu(0)\int^\infty_0 \frac{ds}{s^{ 2}} e^{-\bar M^2_\nu s-
\frac{n^2}{4T^2 s}}= \frac{32 \pi}{T^3} \sum_\nu \psi^2_\nu (0) \bar M_\nu
\frac{ K_1  \left( \frac{n \bar M_\nu}{T}\right)}{n}, \label{a2}\ee where $\bar
M^2_\nu = m^2_f + m^2_\nu$. The $P^{(f)}_q$ acquire the form \be \frac{
P_q^{(f)}}{T^4} = \frac{8 N_c}{\pi T^3} \sum_n \psi^2_\nu (0) \bar M_\nu \sum_n
\frac{(-)^{n+1}}{n} K_1 \left( \frac{n\bar M_\nu}{T}\right) L^n\cosh \frac{\mu
n}{T}.\label{a3}\ee

Now using the representation \be K_1 \left( \frac{n\bar M_\nu}{T}\right)=
 \frac{n\bar M_\nu}{T} \int^\infty_0   e^{ -\frac{n\bar M_\nu}{T} \cosh t} \sinh^2 t\,dt, \label{a4}\ee
  one can sum up the geometrical progression in $n$ with the result
 \be \frac{
P_q^{(f)}}{T^4} = \frac{8 N_c}{\pi T^4} \sum^\infty_{\nu=0}  \psi^2_\nu (0)
\bar M_\nu^2 \int^\infty_0
   \sinh^2 t\,dt\,
 \frac12 \left( \frac{c_+}{1+c_+} + \frac{c_-}{1+c_-}\right),\label{a5}\ee
 where  $c_\pm =\exp \left\{ - \frac{\bar M_\nu}{T} \cosh t -\frac{V_1}{2T} \pm
 \frac{\mu}{T}\right\}.$

 It is interesting to find the exact position and the character of  singularities
 in the complex $\mu$ plane, shown in Fig. \ref{fig:01}. To this end we are writing
 $\frac{\mu}{T}$ in the neighborhood of the point in Fig. \ref{fig:01} with imaginary and
 real parts $i\pi$ and $\frac{\mu_R}{T} $ respectively as
 \be \frac{\mu}{T} =i\pi + \frac{M_\nu + V_1/2}{T} + \frac{ M_\nu}{T}
 y.\label{32'}\ee

 The integral (\ref{a5}) as a function of $y$ is proportional to the function
 $f(y)$,
 \be f(y) = \int^\infty_0 \frac{ t^2 dt F(t)}{t^2 - 2y + O(t^2y, t^4y^2)},
 \label{32''}\ee
  where we have separated the region of small $t$, contributing to the
  singularity, and $F(t\to 0)=$ const.

  One can easily see in (\ref{32''}), that $f(y)$ has a square root singularity
  near $y=0$ and the cut $\operatorname{Re} y\geq 0$, with the discontinuity
  \be f(y+i\delta) - f (y - i\delta) =  i  \sqrt{2y}
  F(\sqrt{2y}).\label{32'''}\ee

  Note, that the branch points $y=0$ occur for every $M_\nu, \nu=0,1,2,...$ and
  this situation is similar to the two-body thresholds in the energy plane with
  ever increasing number of particles.

In the case of the   oscillator-type CM  interaction one has \be \psi^2_\nu (0)
= \frac{M^2_0}{4\pi}, ~~\bar M_\nu^2 = m^2_f + M^2_0 (2\nu+1), ~~ M_0 \approx 2
\sqrt{\sigma_s}.\label{a6}\ee

One can see in (\ref{a5}) that for not large $T$ the lowest in $\nu $ terms
dominate.

In the free limit one has $\sum_\nu \psi^2_\nu(0) \to \int
\frac{d^2p}{(2\pi)^2}$, $\bar M^2_\nu = 2 p^2 + m^2_f$, and for $\Phi_n^{(\rm
free)}$ one obtains from (\ref{a2}) \be\frac{\Phi_n^{(\rm free)} (T)}{n^4} =
\frac{8}{n^4} \int^\infty_0 z^2 K_1 (z) dz =\frac{16}{n^4}.\label{a7}\ee

As a result the limit of no CM interaction is \be \frac{P_q^{(f)} \text{ (no\ CM)}}{T^4} = \frac{4N_c}{\pi^2} \sum^\infty_{n=1} \frac{(-)^{n+1}}{n^4} L^n
\cosh \frac{\mu n }{T}, \label{a8}\ee which reduces to the expression, found in
\cite{27a,27b}

\be \frac{P_q^{(f)} \text{ (no\ CM)}}{T^4} = \frac{1}{\pi^2} \left[ \Psi \left(
\frac{\mu-\frac{V_1}{2}}{T}\right)+\Psi \left(-
\frac{\mu+\frac{V_1}{2}}{T}\right)\right],\label{a9}\ee where \be\Psi(a) = \int
^\infty_0 \frac{z^4dz}{\sqrt{z^2+\nu^2}}\frac{1}{\exp (\sqrt{z^2+\nu^2}-a)+1},
\label{10a}\ee
 and $\nu = \frac{m_q}{T}$.

\section{Numerical calculations and comparison to lattice data}

In this section we present results of calculations for the total pressure

    \be P(\mu,T) = P_{gl} (T) + \sum_{m_q(i)} P^{(f)}_q (\mu_i, T),
    \label{15}\ee
    where $P(\mu_f,T)$ in general depends on the  $\mu_f$ for a given flavor.Below we consider the simplest
    case of equal $\mu_f=\mu$, where $f=u,\,d,\,s$, and the quark masses $m_u=m_d=0$, $m_s=0.1$ GeV.
   $P_{gl}(T)$ is given by Eq.(\ref{1}) and is $\mu$-independent in our approximation of no interaction between quarks and gluons. As for $P_{q} (T,\,\mu)$ we shall use two different strategies for its numerical calculation.

\begin{figure}[htb]
\vspace{0.5cm}
\setlength{\unitlength}{1.0cm}
\begin{center}
\vspace{-0.1cm}
\center{\includegraphics[width=0.8\linewidth]{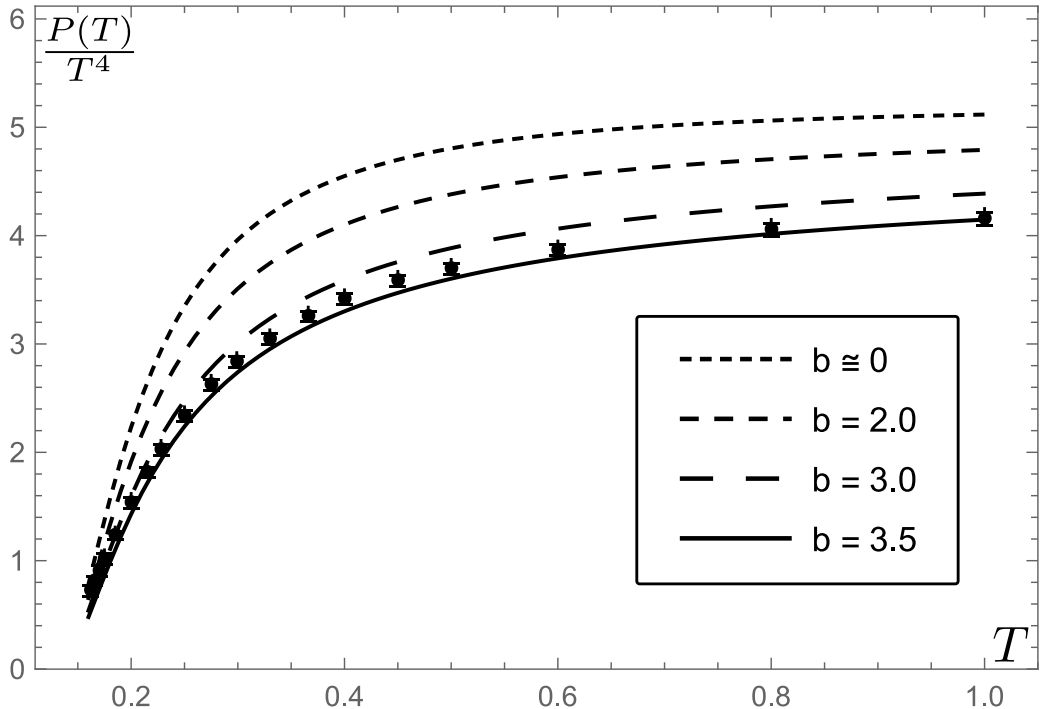}}
\vspace{-0.1cm}
\caption{The pressure $\frac{P(T)}{T^4}$ with $M_0=b\,\sqrt{\sigma_s}$ $(\mu={0})$, where $b={0,\,2.0,\,3.0,\,3.5}$ {(from top to bottom)}, -- filled dots are for the lattice data from \cite{13}.}
\label{fig:02}
\end{center}
\end{figure}

    In the first case one exploits the general form of Eq. (\ref{10}) and approximates the linear confinement case of $S_3^{lin}(s)$ as in Eq. (\ref{9}) with $\bar{M}=\sqrt{m_f^2 + \frac{M_0^2}{4}}$,and $M_0=b\,\sqrt{\sigma_s}$,where $b$ is of the order of 1.
    As a result one obtains $P_q$ as in Eq. (\ref{24}) with $\xi_1$ given in Eq. (\ref{23}).
    The resulting values of $\frac{P(0,T)}{T^4}$ for $\mu=0$ are given in Fig. \ref{fig:02} for $b=0,\,2.0,\,3.0,\,3.5$ in comparison with lattice data from \cite{13}. Note, that for the Polyakov loop $L_f(T)=\exp(-V_1(T)/2T)$ we are using as in Eq. (\ref{1}) and Eq. (\ref{10}) the same values as in \cite{23,27a,27b,28},with $L_{adj}=(L_f)^{9/4}$, namely \be V_1(T)=\dfrac{0.175\,\text{GeV}}{1.35\frac{T}{T_0}-1},\quad T_0=0.16\,\text{GeV} \ee in the interval $0.16 \,\text{GeV}< T < 1 \,\text{GeV}$.

\smallskip

\begin{figure}[htb]
\vspace{0.5cm}
\setlength{\unitlength}{1.0cm}
\begin{center}
\vspace{-0.1cm}
\center{\includegraphics[width=0.8\linewidth]{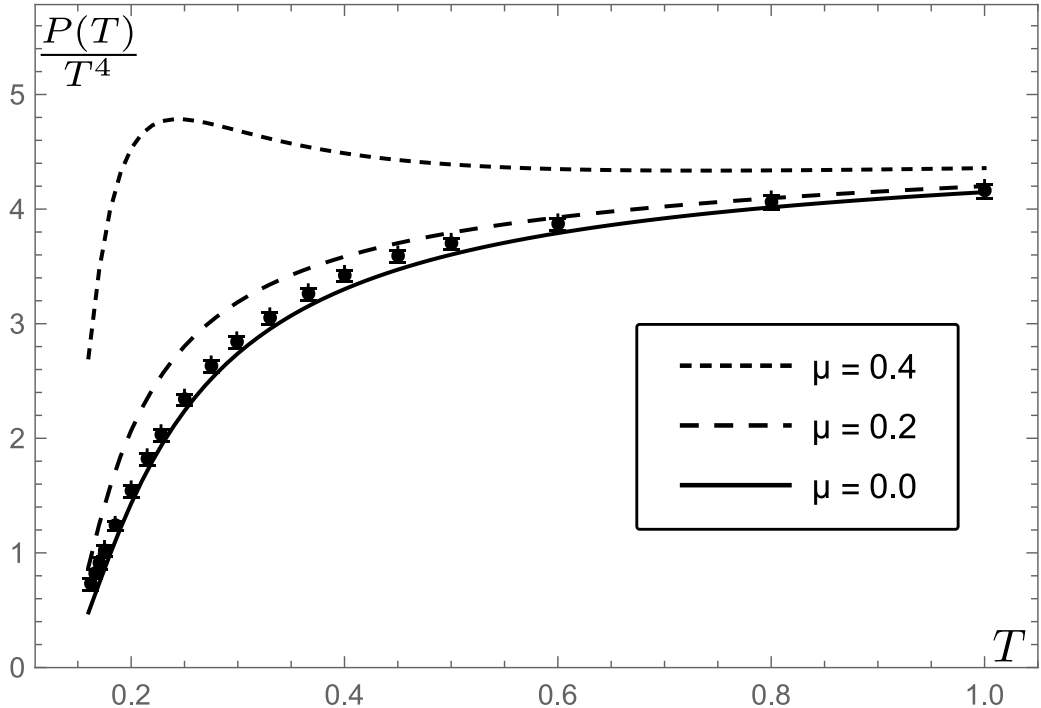}}
\vspace{-0.1cm}
\caption{The pressure $\frac{P(T)}{T^4}$ with $M_0={3.5\,\sqrt{\sigma_s}}$ for $\mu={0.0,\,0.2,\,0.4}$ (from bottom to top), -- filled dots are for the lattice data from \cite{13}.}
\label{fig:03}
\end{center}
\end{figure}

    The resulting values of $L_f(T)$ are in the same domain as the lattice data of \cite{1*,2*,3*,4*} for $T < 0.5 \,\text{GeV}$,while at higher $T$ our $L_f$ is smaller due to necessary renormalization,because of different definitions, see \cite{5*} for the discussion of this point.
    Of special importance is the possible $\mu$-dependence of $L_f(T)$,which can occur due to density dependence of vacuum fields, as well as due to quark-quark interaction.In our approach at this stage we disregard this dependence ,which is partly supported by lattice data \cite{6*}.
    As a result one can see in Fig. \ref{fig:02} a reasonable agreement of our curve for $P(T,\mu=0)$ for $M_0=3.5\,\sqrt{\sigma_s}$ with lattice data from \cite{13}, and we shall exploit this value for $\mu > 0$.
    In Fig. \ref{fig:03} we show the behavior of $P(T,\,\mu)$ for $\mu=0,\,0.2,\,0.4$ GeV. One ca see a maximum appearing at large $\mu$ around $T=0.25$ GeV.

\smallskip

\begin{figure}[htb]
\vspace{0.5cm}
\setlength{\unitlength}{1.0cm}
\begin{center}
\vspace{-0.1cm}
\center{\includegraphics[width=0.8\linewidth]{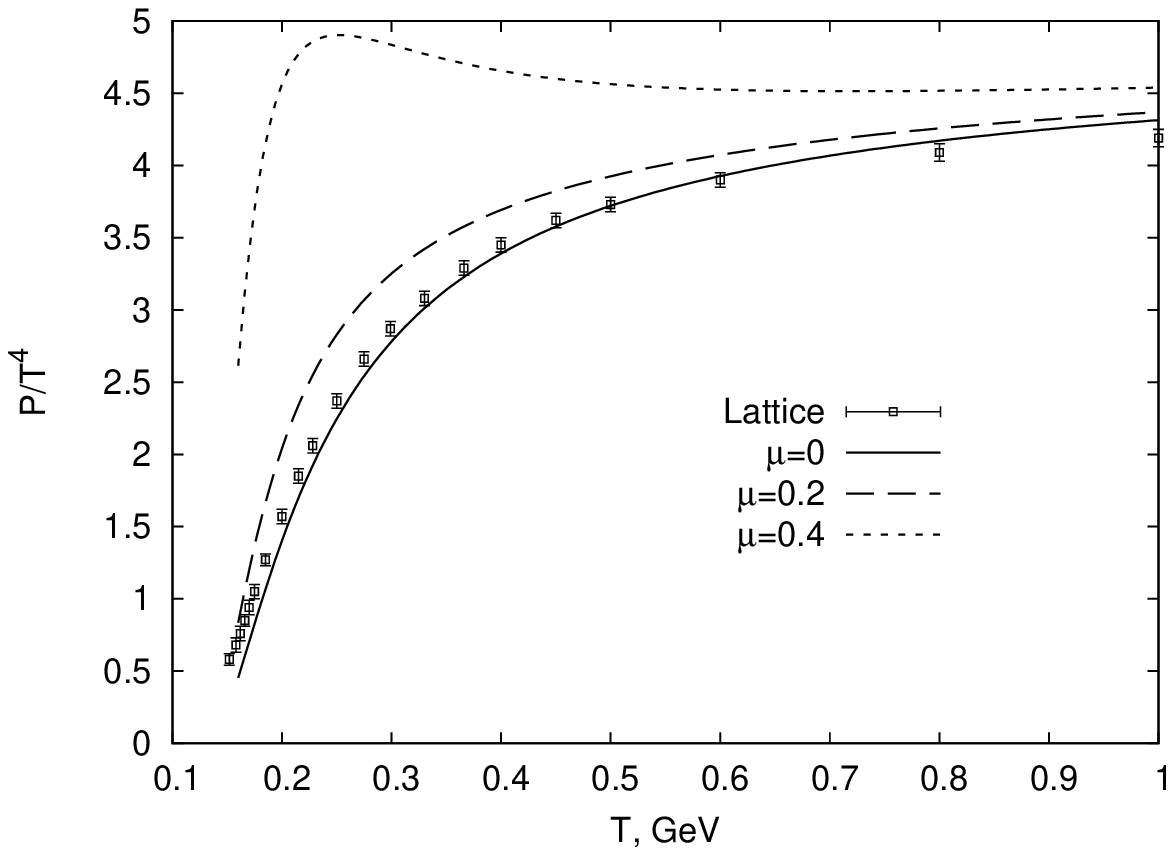}}
\vspace{-0.1cm}
\caption{The pressure $\frac{P(T)}{T^4}$ with $M_0={3.5\,\sqrt{\sigma_s}}$ for $\mu={0.0,\,0.2,\,0.4}$ (from bottom to top), empty squares with error bars are for the lattice data from \cite{13}.}
\label{fig:04}
\end{center}
\end{figure}

    At the same time we are using another strategy, exploiting Eq. (\ref{a3}) for $P_f(T,\,\mu)$ with the values $\psi_{\nu}$ and $M_{\nu}$ from (\ref{a6}), corresponding to the oscillator CM interaction. The resulting behavior of the $P(T,\,\mu)/T^4$ is shown in Fig. \ref{fig:04} for $\mu=0,\,0.2,\,0.4$ GeV, where also the case of $\mu=0$ can be compared to lattice data \cite{13}. One can see a reasonable agreement with lattice data for $\mu=0$ and an agreement with the results of Fig. \ref{fig:03},obtained in the first approach,which can be considered as additional support of our results.

\smallskip

\section{Conclusions and prospectives.}

We have considered above in the paper the propeties of the quark-gluon medium in the temperature interval
$0.15 \,\text{GeV} < T < 1 \,\text{GeV}$ and for the chemical potential $\mu=0,\,0.2$ and $0.4\,\text{GeV}$.
We have taken into account only the \textit{np} part of interaction, which is connected with the vacuum fields and we have disregarded in this first part of study the effects of the continuous phase transition(crossover), which mostly proceed in the lower temperature interval, and will be considered elsewhere.
The main reason for our choice of dynamics is the fact, that the CM confinement and Polyuakov interaction $(V_1(T))$ are most strong in this region and moreover CM confinement is growing with temperature.

As it is, we have analyzed the behavior of $P(T,\,\mu)$ and up to $\mu=0.4\,\text{GeV}$ and have not found any
discontinuous effects in this area. It is seen in Figs. \ref{fig:02}-\ref{fig:04} that the pressure has a smooth behavior, while the
peak in $P(T,\,\mu)$ appears at smaller $T$ with increasing $\mu$. It is important, that the series over $n$ in Eq. (\ref{10}) is convergent for these values of $\mu$, as it was checked both via the sum over eigenstates, Eq. (\ref{a3}), and via the linear approximation, Eq. (\ref{9}).
At the same time we have studied above the analytic properties of thermodynamic potentials in the complex $\mu$
plane and have found sequences of branch points with cuts,going outwards,see Fig. \ref{fig:01}.
These singularities and cuts are dynamically explained by the Polyakov line interaction $V_1(T)$ and CM confinement eigenvalues $M_{\nu}$, as it is shown in Eq. (\ref{A2}), indeed at the branch point one has $\pm\mu=M_{\nu} + \frac{V_1(T)}{2}$.
We have not studied above the explicit consequences of these singularities for the convergence of the series
and the possibility of quasicritical points, leaving this topic for the future.\medskip

\PRLsep

The authors are grateful to V.G.~Bornyakov for useful discussions. This work was done in the framework of the scientific project,supported by the Russian Scientific Fund, grant \#16-12-10414.


\begin{thebibliography}{99}

\bibitem{1}
J.~C. Collins, M.~J. Perry,  Phys. Rev. Lett.  {\bf 34}, 1353 (1975). 

\bibitem{2}
N.~Cabibbo, G.~Parisi, Phys. Lett. {\bf B 59},  67 (1975). 

\bibitem{3}
E.~V. Shuryak,   Sov. Phys. JETP  {\bf 47}, 212 (1978). 

\bibitem{4}
M.~Creutz,  Phys. Rev. {\bf D 21}, 2308 (1980). 

\bibitem{5}
L.~D. McLerran, B.~Svetitsky,  Phys. Lett. {\bf  B 98}, 195 (1981). 

\bibitem{6}
J.~Kuti, J.~Polonyi, K.~Szlachanyi, Phys. Lett. {\bf B 98}, 199 (1981). 

\bibitem{7a}
P.~Braun-Munzinger, V.~Koch, Th.~Sch\"{a}fer and J.~Stachel, Phys.
Rept. {\bf 621}, 76 (2016) [arXiv:1510.00442 [nucl-th]].

\bibitem{7b}
A.~Andronic, P.~Braun-Munzinger, K.~Redlich and J.~Stachel, J. Phys.: Conf. Ser. {\bf 779}, 012012 (2017) [arXiv:1611.01347 [nucl-th]].

\bibitem{8a}
M.~Albright, J.~Kapusta and C.~Young, Phys. Rev. {\bf C 90}, 024915 (2014) [arXiv:1404.7540 [nucl-th]].

\bibitem{8b}
M.~Albright, J.~Kapusta and C.~Young, Phys. Rev. {\bf C 92}, 044904 (2015) [arXiv:1506.03408 [nucl-th]].

\bibitem{9}
J.~Cleymans, H.~Oeschler, K.~Redlich and S.~Wheaton, Phys. Rev. {\bf C 73}, 034905 (2006) [arXiv:hep-ph/0511094].

\bibitem{10}
Sz.~Borsanyi, G.~Endr\"{o}di, Z.~Fodor, A.D.~Katz and K.K.~Szabo, JHEP,
{\bf 2012}(07): 056 (2012), [arXiv:1204.6184 [hep-lat]].

\bibitem{11a}
L.~Giusti and M.~Pepe, Phys. Lett. {\bf B 769}, 385 (2017) [arXiv:1612.00265 [hep-lat]].

\bibitem{11b}
L.~Giusti and M.~Pepe, PoS (LATTICE-2016): 061 [arXiv:1612.02337 [hep-lat]].

\bibitem{12}
A.~Bazavov, T.~Bhattacharya, G.~De Tar, et al, (Hot QCD Collaboration), Phys. Rev. {\bf D 90}, 094503 (2014) [arXiv:1407.6387 [hep-lat]].

\bibitem{13}
Sz.~Borsanyi, G.~Endr\"{o}di, Z.~Fodor, et al., JHEP {\bf 2010}(11): 077  (2010) [arXiv:1007.2580 [hep-lat]].

\bibitem{14}
E.~Braaten and R.D.~Pisarski, {Phys. Rev. Lett.} {\bf 64}, 1338 (1990). 

\bibitem{15}
J.O.~Andersen, E.~Braaten and M.~Strickland,    {Phys. Rev. Lett.} {\bf 83}, 2139 (1999) [arXiv:hep-ph/9902327].

\bibitem{16}
N.~Haque, A.~Bandyopadhyay, J.O.~Andersen, M.G.~Mustafa,  M.~Strickland, et al., JHEP {\bf 2014}(05): 027 (2014) [arXiv:1402.6907 [hep-ph]].

\bibitem{17a}
A.D.~Linde, Phys. Lett. {\bf B 96}, 289 (1980). 

\bibitem{17b}
Yu.A.~Simonov, arXiv:1605.07060 [hep-ph].

\bibitem{18a}
Yu.A.~Simonov, JETP Lett. {\bf 54}, 249 (1991). 

\bibitem{18b}
Yu.A.~Simonov, JETP Lett. {\bf 55}, 627 (1992). 

\bibitem{18c}
Yu.A.~Simonov, Phys. At. Nucl. {\bf 58}, 309 (1995) [hep-ph/9311216].

\bibitem{18d}
N.O.~Agasian, JETP Lett.  {\bf 57}, 208 (1993). 

\bibitem{18e}
N.O.~Agasian, JETP Lett. {\bf 71}, 43 (2000). 

\bibitem{18f}
H.G.~Dosch, H.J.~Pirner, Yu.A.~Simonov, Phys. Lett. {\bf B 349} 335 (1995). 

\bibitem{18g}
Yu.A.~Simonov,  in: \textit{``Varenna-1995: Selected Topics in Nonperturbative QCD''}, Eds. A.Di~Giacomo and D.~Diakonov (IOS Press, Italy, 1996), p. 319 [hep-ph/9509404].

\bibitem{19a}
Yu.A.~Simonov, Ann. Phys. {\bf 323}, 783 (2008) [arXiv: hep-ph/0702266].

\bibitem{19b}
E.V.~Komarov, Yu.A.~Simonov, Ann.  Phys. {\bf 323}, 1230 (2008) [arXiv:0707.0781 [hep-ph]].

\bibitem{20}
N.O.~Agasian,  M.S.~Lukashov  and Yu.A.~Simonov, Mod. Phys. Lett. {\bf A 31}, 1050222 (2016) [arXiv:1610.01472 [hep-lat]].

\bibitem{21}
N.O.~Agasian,  M.S.~Lukashov  and Yu.A.~Simonov, Eur. Phys. J. {\bf A 53}: 138 (2017) [arXiv:1701.07959 [hep-ph]].

\bibitem{22a}
O.~Kaczmarek, F.~Karsch, E.~Laermann and M.~Lutgemeier, Phys. Rev. {\bf D 62} (2000) 034021 [arXiv:hep-lat/9908010].

\bibitem{22b}
P.~Bicudo and N.~Cardoso, Phys. Rev. {\bf D 85}, 077501 (2012) [arXiv:1111.1317 [hep-lat]].

\bibitem{22c}
P.Bicudo and N.Cardoso, arXiv:1608.07742 [hep-lat].

\bibitem{22d}
P.~Cea, L.~Cosmai, F.~Cuteri and A.~Papa, JHEP {\bf 2016}(06): 033 (2016) [arXiv:1511.01783 [hep-lat]].

\bibitem{23}
M.S.~Lukashov and Yu.A.~Simonov, JETP Letters {\bf 105}, 691 (2017) [arXiv: 1703.06666 [hep-ph]].

\bibitem{24}
S.A.~Gottlieb, W.~Lin, D.~Toussaint, R.L.~Reken and R.L.~Sugar, Phys. Rev. Lett. {\bf 59}, 2247 (1987). 

\bibitem{25}
M.~D'Elia, G.~Gagliardi and F.~Sanfilippo, Phys. Rev. {\bf D 95}, 094503 (2017) [arXiv:1611.08285 [hep-lat]].

\bibitem{26}
A.~Bazavov, H.-T.~Ding, P.~Hegde, et al., Phys. Rev. {\bf D 95}, 054504 (2017) [arXiv:1701.04325 [hep-lat]].

\bibitem{27a}
Yu.A.~Simonov, M.A.~Trusov, JETP Lett. {\bf 85}, 598 (2007) [arXiv:hep-ph/0703228].

\bibitem{27b}
Yu.A.~Simonov, M.A.~Trusov, Phys. Lett. {\bf B 650}, 36 (2007) [arXiv:hep-ph/0703277].

\bibitem{28}
A.V.~Nefediev, Yu.A.~Simonov  and M.A.~Trusov, Int. J. Mod. Phys. {\bf E 18}, 549 (2009) [ 	arXiv:0902.0125 [hep-ph]].

\bibitem{29}
C.~Bonati, M.~D'Elia, M.~Mariti, M.~Mesiti and F.~Negro, Proceedings of CPOD-2016, C-16-05-30.7 (2016) [arXiv:1610.03338 [hep-lat]].

\bibitem{30}
A.~Roberge and N.~Weiss, Nucl. Phys. {\bf B 275}, 734 (1986). 

\bibitem{31}
C.~Bonati, P.~De~Forcrand,  M.~D'Elia, O.~Philipsen and F.~Sanfilippo, Phys. Rev. {\bf D 90}, 074030 [arXiv:1408.5086].

\bibitem{32}
R.~Falcone, E.~Laermann and M.P.~Lombardo, PoS (LATTICE-2010): 183 [arXiv:1012.4694 [hep-lat]].

\bibitem{33}
C.~De Tar, L.~Levkova, S.~Gottlieb, et al., Phys. Rev. {\bf D 81}, 114504 (2010) [arXiv:1003.5682 [hep-lat]].

\bibitem{34}
S.~Gupta, N.~Karthik and P.~Majumdar, Phys. Rev. {\bf D 90}, 034001 (2014) [arXiv:1405.2206 [hep-lat]].

\bibitem{1*}
A.~Bazavov, N.~Brambilla, H.-T.~Ding, et al., Phys. Rev. {\bf D 93}, 114502 (2016) [arXiv:1603.06637 [hep-lat]].

\bibitem{2*}
P.~Petreczky, H.-P.~Schadler, Phys. Rev. {\bf D 92}, 094517 (2015) [arXiv:1509.07874 [hep-lat]].

\bibitem{3*}
O.~Kaczmarek, F.~Zantow, Phys. Rev. {\bf D 71}, 114510 (2005) [arXiv:hep-lat/0503017].

\bibitem{4*}
S.~Borsanyi, S.~Durr, Z.~Fodor, et al., JHEP {\bf 2012}(08): 126 (2012) [arXiv:1205.0440 [hep-lat]].

\bibitem{5*}
Yu.A.~Simonov, Phys. Lett. {\bf B 619}, 293 (2005) [hep-ph/0502078].

\bibitem{6*}
J.~Takahashi, K.~Nagata, T.~Saito, et al., Phys. Rev. {\bf D 88}, 114504 (2013) [arXiv:1308.2489 [hep-lat]].

\end{thebibliography}
\end{document}